\documentclass[sigconf, nonacm=True]{acmart}

\usepackage{subcaption}
\usepackage[ruled,vlined,linesnumbered]{algorithm2e}
\usepackage{amsmath}
\usepackage{mathtools}
\usepackage{bbm}
\usepackage{color}
\usepackage{balance}
\usepackage{courier}

\begin{document}

\title{Heterogeneous Effects of Software Patches in a\\Multiplayer Online Battle Arena Game}


\author[Y. He]{Yuzi He} 
\affiliation{
    \department{Department of Physics \& Astronomy}
    \department{Information Sciences Institute}
    \institution{University of Southern California}
    \country{United States}
}
\email{yuzihe@usc.edu}
\author[C. Tran]{Christopher Tran} 
\affiliation{
    \department{Department of Computer Science}
    \institution{University of Illinois at Chicago }
    \country{United States}
}
\email{ctran29@uic.edu}
\author[J. Jiang]{Julie Jiang} 
\affiliation{
    \department{Department of Computer Science}
    \department{Information Sciences Institute}
    \institution{University of Southern California}
    \country{United States}
}
\author[K. Burghardt]{Keith Burghardt}
\affiliation{
    \department{Information Sciences Institute}
    \institution{University of Southern California}
    \country{United States}
}
\email{keithab@isi.edu}
\author[E. Ferrara]{Emilio Ferrara} 
\affiliation{
    \department{Department of Computer Science}
    \department{Information Sciences Institute}
    \institution{University of Southern California}
    \country{United States}
}
\author[E. Zheleva]{Elena Zheleva} 
\affiliation{
    \department{Department of Computer Science}
    \institution{University of Illinois at Chicago }
    \country{United States}
}
\author[K. Lerman]{Kristina Lerman}
\affiliation{
    \department{Information Sciences Institute}
    \institution{University of Southern California}
    \country{United States}
}
\email{lerman@isi.edu}


\begin{abstract}
    The popularity of online gaming has grown dramatically, driven in part by streaming and the billion-dollar e-sports industry. Online games regularly update their software to fix bugs, add functionality that improve the game's look and feel, and change the game mechanics to keep the games fun and challenging. 
    An open question, however, is the impact of these changes on player performance and game balance, as well as how players adapt to these sudden changes. 
    To address these questions, we use causal inference to measure the impact of 
    software patches to League of Legends, a popular team-based multiplayer online game. We show that game patches have substantially different impacts on players depending on their skill level and whether they take breaks between games. We find that the gap between good and bad players \emph{increases} after a patch, despite efforts to make gameplay more equal. Moreover, longer between-game breaks tend to improve player performance after patches. Overall, our results highlight the utility of causal inference, and specifically heterogeneous treatment effect estimation, as a tool to quantify the complex mechanisms of game balance and its interplay with players' performance.
\end{abstract}

\maketitle

\section{Introduction}
Interest in online gaming has grown explosively, partly due to the burgeoning e-sports industry and live streaming technologies that enable people to watch others play in real time.  League of Legends (LoL) is one of the most popular  \textit{multiplayer online battle arena} (MOBA) games with tens of millions of monthly active players. They compete in teams (typically of five  players) to capture the opposing team's base. In a LoL match, each player controls a character---known as  \textit{champion}---and uses it to plan attacks or mount defenses against opponents. Champions vary in their power levels and abilities (type of spells cast, rate of attack, armor strength, and the like), which can be further enhanced by the player during a match by means of skill points gained or items earned. Champions also differ in play styles, some being devoted to cure teammates, others to effectively defeat opponents, etc., leading to different \textit{classes of champions}.

Likewise, individual players differ in their skill level, game style, and mastery of specific champions or champion classes. MOBA games typically  match players together into teams to balance the teams' skills, so that neither side will have a built-in advantage in winning the match. Game balance makes gameplay more fun and engaging for the players~\cite{altimira2017enhancing}.

Riot Games, the developer of LoL, regularly updates the game by releasing new versions of the game's software, known as \textit{software patches}. These patches not only fix bugs and make technology improvements, but also introduce new functionality, content, and game balance. One study of the first six seasons of LoL identified over 7,000 changes made in 164 software patches~\cite{claypool2017impact}. Although users will first notice cosmetic changes to the look and feel of the game (e.g., changing the champions' graphical appearance, a.k.a. skins), many of the patches affect gameplay by modifying the abilities of champions. These types of changes can be classified as \textit{buffs} that increase a champion's strength (e.g., bulking up their armor), \textit{nerfs}, which are more common, that decrease strength (e.g., reducing the distance of their spells), or neutral changes that do not substantially impact champion abilities. Patches are often created following the introduction of new features that gave too much power to some champions or skills. When this happens, the game balance is altered, making the game too difficult---or too easy---for players and, therefore, less fun and engaging. To restore game balance, MOBA games are regularly patched to nerf to overpowered champions and buff underpowered ones. 

Measuring the impact of software patches on player performance and game balance is difficult due the to complex interplay between the choices of players and game outcomes. A nerfed champion, when played in combination with other champions, may improve a team's chances of winning. Player characteristics and play styles may also affect its effectiveness. These considerations dramatically complicate the maintenance of game balance, especially as new champions and skills are regularly released to expand game features and keep it interesting for the players. Surprisingly, there has been relatively little work done on this problem. Existing research examined the impact software patches on player's choices, showing that they increase player preference for buffed champions and reduce preference for nerfed champions~\cite{wang2020research}. Additionally, buffs typically improve champion's win rate, especially for underperforming champions~\cite{claypool2017impact} 


When measuring the impact of an intervention (e.g., a patch) on a system, it is important to model the intervention's causal effect on target outcomes rather than the correlation between them~\cite{pearl-book09}. 
This is especially true in \emph{observational studies}, since we never observe all potential outcomes (this is the fundamental problem of causal inference).
Additionally, the specification of a treatment (software patch) is typically biased, either through confounding or selection bias~\cite{bareinboim-pnas16}.
Not all champions are selected to be buffed or nerfed in a single patch (confounding), and not all players play the same champions (selection bias). 
Since the distribution of samples may differ between treated and control populations, a supervised model naively trained to minimize factual errors would overfit to properties of one group and not generalize well to the population.

We address this problem and extend the state of the art by treating the matches that take place before and immediately after a patch is released as control and treated populations, respectively (users are as-if randomly selected to each group). 
We then use causal inference methods to measure the effect of the patches on player performance (the number of kills) and team performance (probability of winning).  
We find a large variation in the average effect of patches across the population. 
The impact of patches, however, also depends on individual player features or champions played. To account for this, we also estimate the \emph{heterogeneous} treatment effect (HTE) of the patch. HTEs measure the effect for different subgroups of a population for which effects differ~\cite{athey-annals19,tran-aaai19,athey-pnas16}. For example, good players may be less affected by a patch than bad players, or a nerf on a champion may not affect the team's win probability if a synergistic champion is still strong. 


We discover that some software patches, for example patch 4.20 and 6.9, can substantially affect player performance. After applying a causal tree HTE model to each patch, we find significant heterogeneity in team performance changes despite LoL's 
game balancing mechanism. Moreover, the effect of patches on player performance varies significantly with the champion type they play and 
their initial performance. Despite the heterogeneity in players and patches, we find some results that generalize across patches: players who take significant breaks between matches perform especially well after a patch. 
In addition, several performance metrics show the significant advantages that patches bring to high-performing players over low-performing ones. Therefore, surprisingly, these patches caused a widening in the gap between the high and low-performance players, which is contrary to the spirit of patching aimed at game balancing. 
Overall, our results underscore the importance of player heterogeneity in policy changes, and limitations of attempts to balance player performance, possibly because player heterogeneity is not taken sufficiently into account. 

\section{Related works}
The rise in popularity of MOBA games in recent years has given researches a wealth of large-scale user-centric datasets. The team-based nature of such games led to many research in optimal team compositions, including identifying and predicting the influence of teammates using co-play networks \cite{SapienzaTeamComposition} and building recommender systems for the line-up of heroes (the equivalent of champions in LoL) in DOTA 2 \cite{hanke2017recommender, chen2018art}, another popular MOBA game. Further, MOBA games boast metagaming strategies, which are collectively decided by players (the crowd) as the most optimal strategy for the team or for each champion. For example, \cite{lee2017identifying} finds that the mostly widely successful team composition in LoL consists of one player in each of the five positions, although some non-meta teams have significant advantages. 

One of the most significant factors influencing gameplay are patches. Patches are regular updates to the game that fix bugs, introduce new game contents, and most importantly alter the skills and abilities of champions to balance the game \cite{claypool2017impact}.
\citeauthor{wang2020research} show that the effect of champion balancing patches also affect player's champion preference \cite{wang2020research}. 
Other game related work focus on mining and understanding human behavior. To be specific, \citeauthor{sapienza2018individual} finds that prolonged gameplaying sessions lead to performance decline, a phenomenon that may be due to cognitive depletion \cite{sapienza2018individual}. \citeauthor{sapienza2018non}, on the other hand, used unsupervised tensor factorization methods to cluster different types of users based on their in-game performance metrics \cite{sapienza2018non}.

The estimation of HTE is an important problem in many fields, even if has not often been applied to games. HTE estimation refers to finding subsets of the population for which causal effects of a treatment differ from the population and other distinct subsets.
Medical professionals, for example, may be interested in how a drug treatment may benefit one group, but potentially have adverse reactions to another group~\cite{shalit-icml17}.
Marketers, alternatively, may be interested in how an advertisement influences different users to buy a product~\cite{bouttou-jmlr13}. 

Many supervised techniques have been developed for HTE estimation~\cite{athey-pnas16,grimmer-pol17, shalit-icml17,xie-kdd18,tian-jasa14, tran-aaai19} and the related problem of finding individualized treatment regimes~\cite{almardini-adv15,laber-bio15,kallus-icml17}. 
Many methods build upon interpretable tree-based methods, such as decision lists~\cite{lakkaraju-aistats17}, 
classification and regression trees (CART)~\cite{athey-pnas16, tran-aaai19, zeileis-jcgs08}, and random forests~\cite{athey-annals19,wager-jasa17}. 
Others follow more in line with the supervised machine learning paradigm, such as using supervised base learners, called meta-learners, which decompose HTE estimation into multiple regression or classification problems~\cite{kunzel-pnas19,nie-rlearner17}. 
Representation learning using deep neural networks have also been proposed for estimating HTEs~\cite{shalit-icml17,johansson-icml16}.

Our paper differs from these previous research due to our focus on how patches heterogeneously affect player and team performance. For example, we test the hypothesis that a good player may be more robust to changes but a bad player's performance may vary substantially with each patch. Additionally, a buffed champion may have an overall increase in winning rate, but if their team includes a nerfed champion, it may actually perform worse in those games.



\section{Background and Game Data}

LoL is a popular MOBA game where two teams of players compete in a match to destroy the opposite team's home base, or \textit{Nexus}. A match lasts about 30 minutes, and is composed of two teams of five players or two teams of three players, depending on the game map selected. At the beginning of every match, each player selects a champion to play as and assume one of the five positions available on the map: Top, Middle, Jungle, Attack Damage Carry, and Support. The dataset tracks individual player's performance in matches, measured by metrics such as the number of \textit{kills} and \textit{assists} the player makes in the match, as well as whether the team for which the player was playing for won the match. At the end of the match, one and only one team will emerge as the winner. 

There are over 130 different champions, each with different powers and abilities. These champions belong to seven disjoint \textit{champion types}: controllers, fighters, mages, marksmen, slayers, tanks and unique playstyles, sorted in alphabetic order. Not every champion is equally popular, and many champions are rarely picked. The most popular champion is \textit{Thresh}, who is chosen by about 3\% of all players, followed by \textit{Lucian} and \textit{Vayne}. In our analysis, we consider the top 25 most popular champions, and these appear in most matches within our dataset.

An LoL season typically starts the beginning of each calendar year, and concludes in late November or early December. Players are ranked in tiers at the end of each season. In between seasons are pre-seasons, in which developers typically introduce large overhauls to the games in preparation for the new season. Games during pre-seasons are not counted towards players' seasonal rankings.

\subsection{Features} 
The LoL dataset was collected from mid-2014 to the end of 2016 \cite{sapienza2018individualData}. It consists of 1.2 million unique players in 437 thousand matches. The dataset contains information about players and matches at different levels of granularity and for different versions of the game. Features granularity in the data is at the match level, user level and season level. At a match level, the data features encode basic information about the match, including the match duration, start time of the match, map ID, queue type, patch ID, season ID and the outcome the match (which team loses and which team wins). The queue type determines the type of the game. For example, a game can be in ranked queue or unranked queue, where a ranked queue indicates that the game outcome contributes to each players final seasonal ranking. The queue type also indicates whether it is a solo queue game, in which the teams are formed by individual players who likely did not know each other beforehand.

At the user level, the dataset records the champion, the role and the lane selected by each user for each match. Every champion belongs to the seven established champion types. Individual in-game performance metrics include  \textit{kills}, \textit{deaths}, \textit{assists}, \textit{gold earned} and \textit{gold spent}, and the champion level achieved by the player in this game (\textit{champLevel}). 

To  track the prior experience of players, we compile a set of user features. For every user-match pair, we track the number of matches the user played thus far and time interval since previous match (\textit{timeSinceLastMatch}). Motivated by research showing evidence that continuous LoL gameplay leads to performance deterioration \cite{sapienza2018individual}, we additionally compute per-session statistics, where a session is defined as a series of matches without a break of at least 15 minutes between consecutive matches. For each user and match, we record the user's session number and the index of the match within this session. Past user behavior includes cumulative and average performance metrics from  user's first match up until (but not including) the current match for \textit{kills}, \textit{deaths}, \textit{assists}, \textit{KDA}, \textit{gold earned}, and \textit{gold spent}, (\textit{mean*AtStart} and \textit{cum*AtStart}).  Per-session aggregated performance metrics calculated in the same fashion (\textit{sessionMean*AtStart} and \textit{sessionCum*AtStart}) 

Finally, at the season level, the features track the highest tier achieved by the player in the previous season (\textit{highestAchievedSeasonTier}), which is only available for players who have played competitively in the previous season. 

\subsection{Game Patches}
Developers release software patches to fix bugs and security vulnerabilities, as well as to release new content and adjust game balance. 
There are 62 software patches in our data set, corresponding to different versions of the game, ranging from version 4.6 to 6.22. We study how buffs and nerfs introduced by a new version of game software impact performance of teams and individual players, and how the impact differs depending on team composition, player characteristics, and game settings.

\section{Methods}

\subsection{Heterogeneous Treatment Effects}

The goal of our study is to find how the effect of software patches differs between different subgroups of players and champions chosen by the team. To discover these effects from data, we consider the problem of heterogeneous treatment effect (HTE) estimation, where the treatment is a new software patch. 
We consider our unit of interest as a match of LoL and the outcome is some result of the match, such as the number of \emph{kills} at the individual player level or \emph{win or loss} at the team level.

Formally, we frame our problem using the Rubin framework of potential outcomes~\cite{rubin-psych74}. Let the treatment for a match $i$ be defined as $W_i$, such there exists potential outcomes $Y_i(W_i =w)$, which is the outcome of a match $i$ on software patch $w$. 
Each match has a set of characteristics (features), $X_i$, such as the champions played or player statistics before the match. 
For any match, we can only observe the outcome from one of the treatments (the patch it was played on) and not at any point in time, defined as $Y_i(W_i=w_i)$. 
The goal of our work is to estimate the HTE of the treatment, which we take to be a software patch, which the game company uses to update the game to a new version. 
Let any two consecutive software patches be versions $w_t$ and $w_{t+1}$. 
The goal of HTE estimation is to estimate the conditional average treatment effect (CATE):
\begin{equation}
    \tau(x_i) = E[Y_i(w_{t+1}) - Y_i(w_{t}) \mid X_i = x_i],
\end{equation}
where $X_i = x_i$ represents a subset of the population. 
In this case, we estimate how an outcome of interest changes between two versions of the game.
For example, we may want to know how \emph{kills} (outcome) of a {player} on a specific \emph{champion} (features) changes from one \emph{version} (treatment) to another.

\begin{figure*}[!htb]
    \centering
    \includegraphics[width=1.0\linewidth]{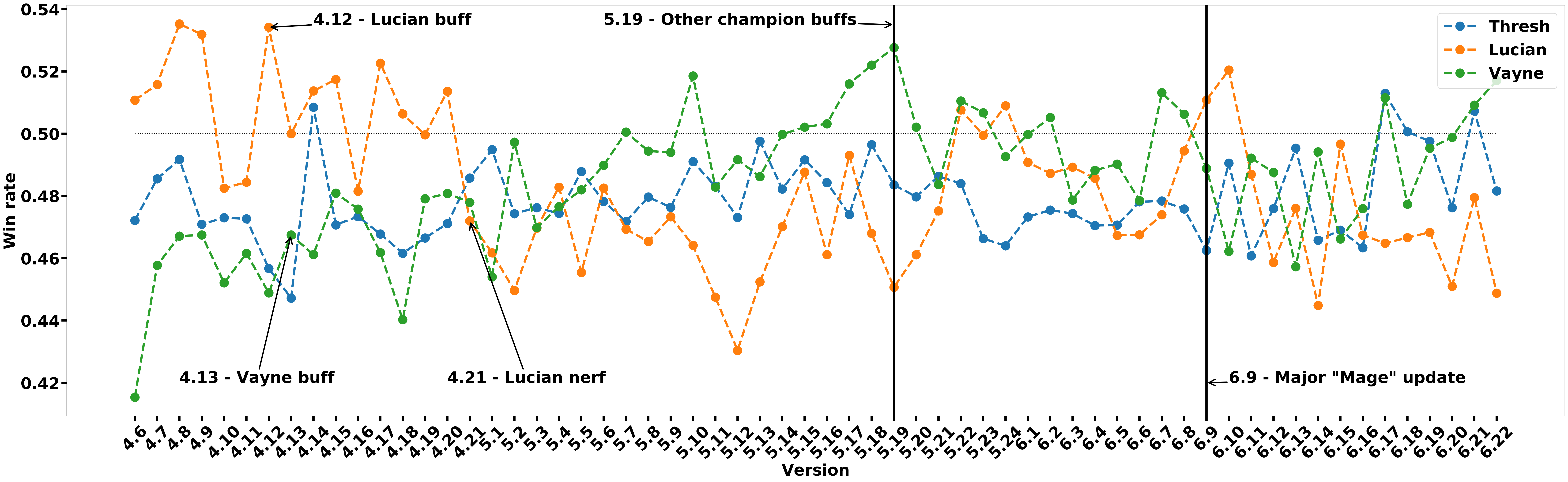}
    \caption{Win rate (\%) of the top-three most played champions for each patch. 
    }
    \label{fig:win_rates}
\end{figure*}

\subsection{Causal Trees for HTEs}

Tree-based methods have been popularized recently for  heterogeneous treatment effect (HTE) estimation~\cite{athey-pnas16,tran-aaai19,wager-jasa17}. 
Causal trees work similarly to CARTs, in that they both greedily partition the feature space based on a certain criterion.
The crucial difference is that a causal tree splits the feature space in order to reduce the expected variance of the estimated HTE while a CART does so in order to minimize the classification or regression loss.
The splitting continues till the predefined minimum leaf node size is reached or there is no more statistically significant splitting can be made.
We use a variant of causal trees developed by Tran and Zheleva~\cite{tran-aaai19}, which introduce a type of validation set for generalizing causal effects to unseen data. 
For our experiments, since the validation set is randomly selected, the causal tree built for each training/validation split will be slightly different. But the overall structures of the causal trees are similar and we did not observe any contradictory results. 
After a causal tree is built, HTE will be estimated for every leaf node of the tree. Given feature $x_i$, the HTE $\tau(x_i)$ is inferred as the HTE for the leaf node corresponds to $x_i$. 

Figure~\ref{fig:win_rate_4.12} shows an example causal tree built using the algorithm developed in~\cite{tran-aaai19}, where the treatment is patch 4.12, and the outcome of interest is whether a team wins or loses. 
In each node, we have an estimated causal effect, the p-value of that effect based on an independent t-test, and how many samples are used to estimate that effect. 
The estimated effect at any node is the difference in means when treated and not treated (e.g. percentage of wins on patch 4.12 compared to patch 4.11). 
At any parent node, there is a splitting feature. 
At the root node, the split is based on the binary feature whether the champion Lucian is on the team or not, if yes then we traverse left, otherwise we traverse right. 
Statistically significant nodes are highlighted by a purple box. 
We discuss these findings in Section~\ref{sec:results}.

\section{Results}
\label{sec:results}

Software patches can affect how champions perform on a global level, and they can also affect how players perform using those champions. 
We first explore how team's probabilities of winning vary by champions chosen by team members, independent of the players playing them. 
We then explore how individual players are affected by software patches.

\subsection{Effect on Team Performance}

In our analysis we focus on the 5 versus 5 play mode and ranked (competitive) matches, as opposed to normal (casual) queues.
The outcome of interest is whether a team wins or not, and the treatment is the patch of interest (e.g., patch 4.20). 
Since the likelihood a particular champion wins the match (its win rate)  varies by patch, 
some combinations of champions are more likely to result in a team win compared to others in different patches.

\subsubsection{Champion win rate}
Figure~\ref{fig:win_rates} shows the overall win rate for the three most played champions: Thresh, Lucian, and Vayne. The win rate is defined as the fraction of matches won by a team playing that champion. The win rate varies by both patch and champion. For example, in patch 4.6, Lucian has a win rate over 50\%, while Vayne and Thresh have a win rate below 50\%. Over time, the win rates of Lucian and Vayne shift above and below 50\% depending on the patch, while Thresh has a more stable win rate slightly below 50\%. Lucian was buffed in patch 4.12, for example, which leads to a higher win rate in matches played after the patch was introduced. In patch 4.21, however, Lucian receives a nerf, which results in a drop in win rate in the next two patches.
Another interesting observation is that the win rates of Lucian and Vayne are negatively correlated ($r=-0.46, p=0.0002$).
This is because 
Lucian and Vayne are typically played in the same \emph{lane} and \emph{role}.
If one champion is stronger then the other champion is more likely to lose in a match up.
Therefore, even a change in a patch not related to Lucian can affect the final performance, which motivates the study of heterogeneity in win rates. 
As an example, Vayne is buffed in 4.13, which increases her win rate slightly, but lowers Lucian's win rate slightly as well. This applies to other champions that are not in the same lane or role. In 5.19, many other champions are buffed which lowers Lucian's win rate, and increases Vayne's win rate. This could be because Vayne is good against the buffed champions, while Lucian is bad against them. 
The opposite could be true in patch 6.9, where there is a major \emph{mage} update.

\begin{figure*}[!htb]
    \begin{subfigure}[b]{.48\linewidth}
        \centering
        \includegraphics[width=1.0\linewidth]{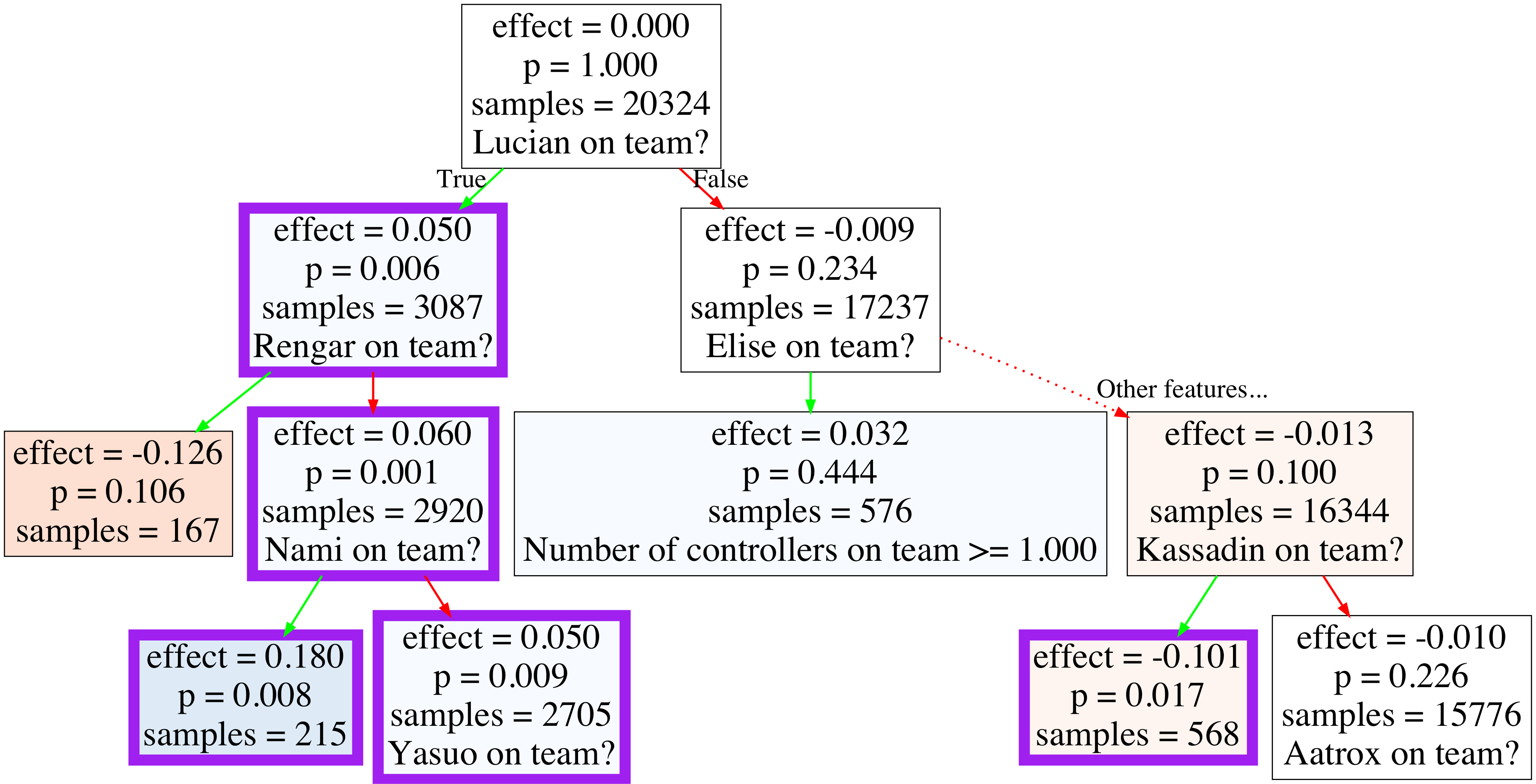}  
        \caption{Trimmed causal tree for patch 4.12}
        \label{fig:win_rate_4.12}
    \end{subfigure}
    \begin{subfigure}[b]{.48\linewidth}
        \centering
        \includegraphics[width=1.0\linewidth]{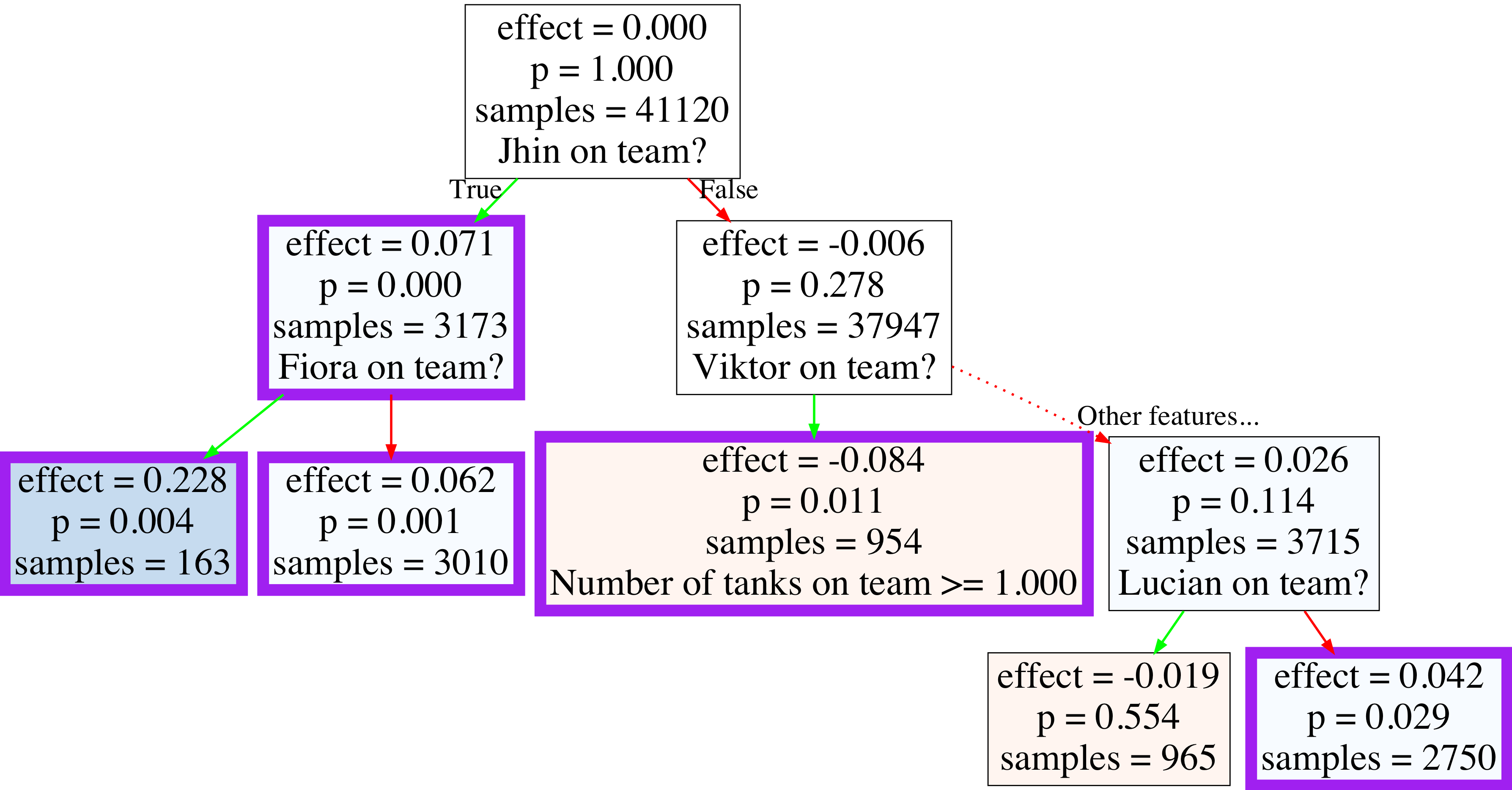}  
        \caption{Trimmed causal tree for patch 6.4}
        \label{fig:win_rate_6.4}
    \end{subfigure}
    \caption{Trimmed causal trees for patch 4.12 (left) and 6.4 (right) which contain Lucian buffs and nerfs, respectively. In the figures, \textit{samples} indicates the number of observations in that node. For non-leaf nodes, it is the total number of observations before splitting.
    }
    \label{fig:win_rate_ct}
\end{figure*}

\subsubsection{Heterogeneous effect on win rate}

\begin{figure*}[!htb]
    \begin{minipage}{0.8\linewidth}
        \centering
        \includegraphics[width=1.0\linewidth]{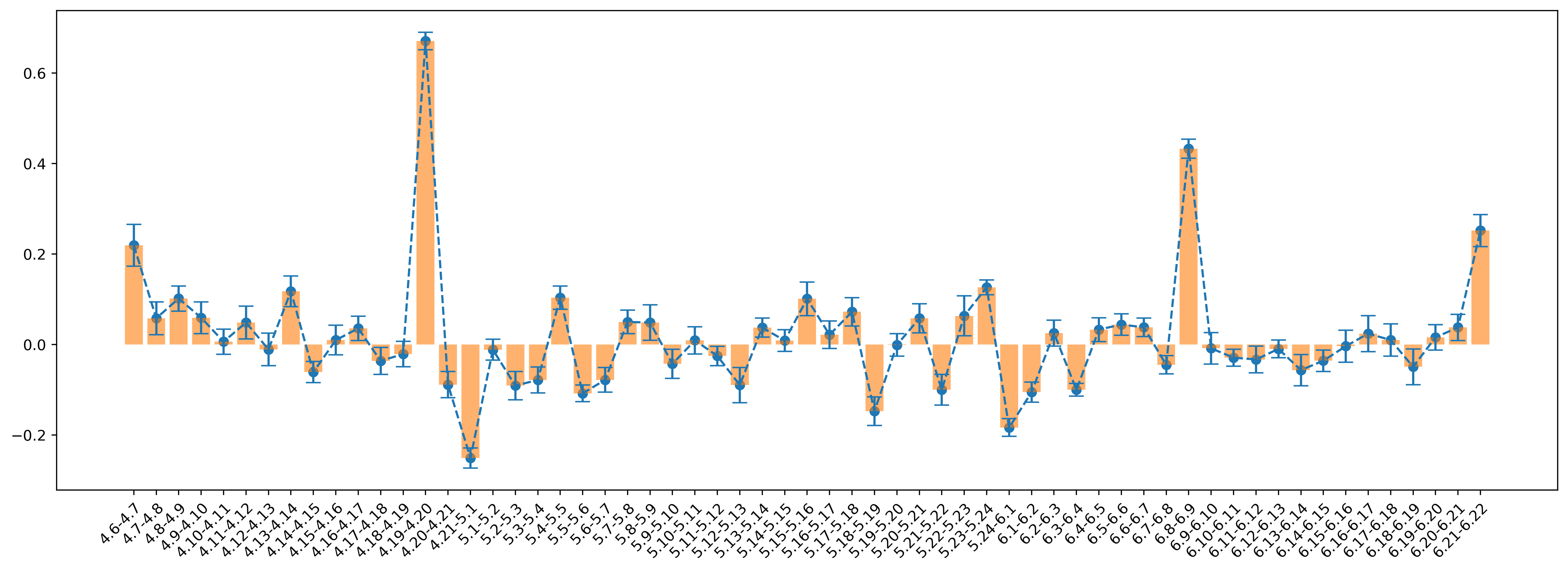}
    \end{minipage}
    \caption{The mean effect of software patches for \textit{kills}. 
 We see there are sharp peaks at patch 4.20 and 6.9.
    }
    \label{fig:version_bar}
\end{figure*}

Next we zoom in on specific patches to see how other features, including champions and types, moderate the effect of the patch on win rate. 
We focus on potential changes on Lucian, since he is one of the three most popular champions, and his win rate varies more than Vayne's and Thresh's.
We choose two patches that contain buffs and nerfs to Lucian, patch 4.12 and 6.4, respectively~\cite{claypool2017impact}.
In addition, we identify two other patches with significant changes to other champions other than Lucian.
The, we build causal trees for these four patches:
\begin{itemize}
    \item 4.12: Significant changes to Lucian, resulting in a strong buff.
    \item 6.4: Small change to Lucian, resulting in a small nerf.
    \item 4.20: Small changes to many champions and largest average increase total kills, which is explored in Section~\ref{sec:player} and shown in Figure~\ref{fig:version_bar}. This is also a ``pre-season'' patch.
    \item 6.9: No changes to Lucian, but major updates  to \emph{mage} champions. This is also a ``mid-season'' patch.
\end{itemize}

We first look at two patches that directly affect Lucian: 4.12 and 6.4.
Figure~\ref{fig:win_rate_ct} shows two trimmed causal trees on this patch, where the omitted features are denoted by a dotted line. The tag ``Other features...'' means we skip to a lower part of the tree.
Nodes with feature splits that have children have been trimmed to save space (e.g. ``Number of controllers on team'' in Figure~\ref{fig:win_rate_4.12}).
This happens in the right side of the trees. We consider the treated population as all matches played after the patch, and the control population as all matches played before the patch. The outcome is team win or loss.

\paragraph{Patch 4.12}
In Figure~\ref{fig:win_rate_4.12}, we see that Lucian's presence on the team is the first splitting feature.
This makes sense as Lucian arguably received the most changes, largely buffs.
This patch increases Lucian's win rate by 5\% as 
shown in the first left node.
For the games where the team included Lucian, if Rengar was also present on the same team, the win rate actually \emph{decreases}  by 12.6\%
, although this effect is not significant with an independent t-test ($p=0.11$). 
In patch 4.12, Rengar received a bug fix that actually results in a nerf, which is acknowledged in the patch notes as a potential nerf: ``This bug fix might be a significant hit to Rengar's jungling effectiveness (he was proccing Madred's Razors twice), but we'll track his performance over time.''
As we traverse further left, if Nami is also on the team, the win rate significantly increases by 18\%.
Although Nami was not changed, likely the combination of Lucian and Nami is significantly improved from the buffs to Lucian.
Further down the right subtree we see that if a team has Kassadin, then it is more likely to lose on patch 4.12.

Importantly,  results from causal analysis \emph{are not} the same as computing the change in win rates individually for each champion. 
For example, Nami's win rate increases by 4.2\% in patch 4.12 without conditioning on other champions and is not significant ($p=0.10$). With Lucian \emph{and} without Rengar, her win rate increases by 23\%, a significant difference. Kassadin has a noticeable nerf in this patch, resulting in an individual win rate decrease of 4.5\% ($p=0.23$), but after removing potential champions in the team, the win rate drops to $10\%$.

\paragraph{Patch 6.4}
Figure~\ref{fig:win_rate_6.4} shows the causal tree for the treatment of patch 6.4. Here, the first split considers whether Jhin is on the team. This is because Jhin receives significant buffs in patch 6.4, which increase his win rate significantly. Traversing the left subtree, if the team has Fiora in addition to Jhin, the win rate goes up significantly by 23\%. 
This observation is interesting because Fiora receives a nerf in patch 6.4.
There could be several reasons, such as the nerf not being enough, Jhin and Fiora being a strong combination, or other champions were changed enough so that Fiora still wins more often than not.
On the right subtree of Figure~\ref{fig:win_rate_6.4}, Viktor and Lucian both have a decrease in win rate, but Lucian's change is not significant. 
Both Viktor and Lucian receive small nerfs.

\paragraph{Patch 4.20}
Patch 4.20 is a ``pre-season'' (season 5) patch.
Generally, pre-season patches contain many more changes than patches released during the season. 
An interesting observation from the tree is that many features selected are champions not directly changed.
At the root node, the first feature selected is whether Jinx is on the team, but Jinx was not changed in patch 4.20. Additionally, several features are based on the champion \emph{types}: the number of \emph{fighters}, \emph{marksmen}, and \emph{mages} affects the win chance of a team.
This effect is heterogeneous: having at least one \emph{fighter} with Jinx increases the win rate by 6.4\%, while not having Jinx and having more than 2 \emph{marksmen} decreases the win rate by 15.4\%.
Further down the tree, we see that our focus champion, Lucian, also has a positive change in win rate if there are at least 2 \emph{mages} on the team.


\begin{figure*}[!htb]
    \begin{minipage}{1.0\linewidth}
        \centering
    Mean performance of players disaggregated by feature \textit{timeSinceLastMatch} \\
        \includegraphics[width=1.0\linewidth]{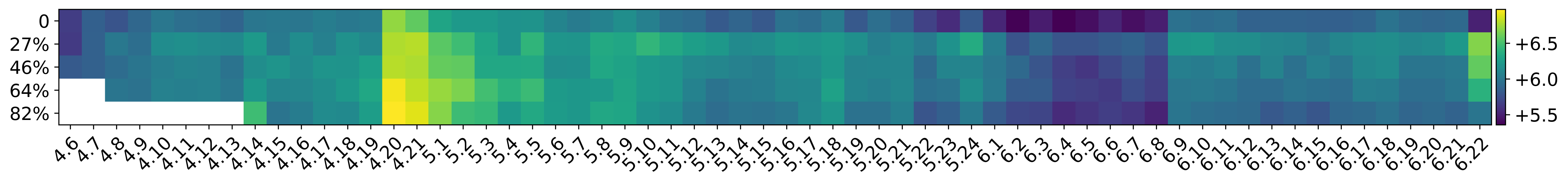}
    \end{minipage}
    \begin{minipage}{1.0\linewidth}
        \centering
    Mean performance of players disaggregated by feature \textit{meanKillsAtStart} \\
        \includegraphics[width=1.0\linewidth]{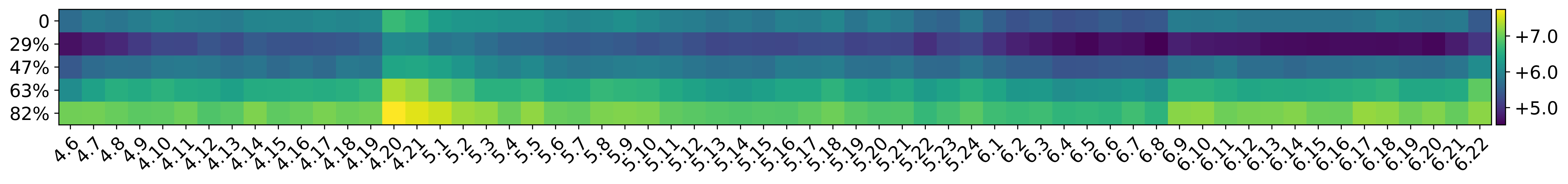}
    \end{minipage}
    \caption{Impact of software patches on player performance. Heatmap shows average player performance, as measured by the number of \textit{kills} per match, for different versions of the game. The upper plot shows the performance of players with different rest time(\textit{timeSinceLastMatch}). In the lower plot, players are binned based on the feature \textit{meanKillsAtStart}, a proxy of player skill. The abrupt color change for versions 4.20 and 6.9 indicates a large difference in performance after the version change.}
    \label{fig:version_heat}
\end{figure*}

\paragraph{Patch 6.9}
This patch is interesting since it changes a significant number of champions type \emph{mages}, which we refer to as the ``major mage update'' in Figure~\ref{fig:win_rates}.
The causal tree learned on patch 6.9 data does not identify many \emph{mages} as features in the top of the tree, except for Brand. On the left subtree there is a split on the number of \emph{controllers}, another champion type similar to mages, which were changed and categorized under the umbrella term of \emph{mages}, which increases the win rate.
Further down the right subtree, Lucian appears as another positive increase in win rate, of 17\% if there is a Riven and no Brand and Wukong.
This explains the overall increase in win rate for Lucian individually shown on patch 6.9 in Figure~\ref{fig:win_rates}.
A potential explanation is the ``major mage update'' increases the amount of \emph{mages} played and Lucian is one \emph{marksman} that may perform better against \emph{mages}.

\paragraph{Summary} Our results demonstrate the nuances of game balance: changes to champions affect the win rate of completely different champions played by the team. 
The computational framework of heterogeneous effect estimation described here can quantify how changes in the abilities of champions reverberate through other champions. Big changes, like major buffs to our focal champion Lucian, are identified early in the tree in patch 4.12 (Figure~\ref{fig:win_rate_4.12}), but small nerfs are identified near the bottom of the tree in patch 6.4 (Figure~\ref{fig:win_rate_6.4}).
Other changes to champions can also affect win rates differently, and even champions who do not change can be affected, such as Jinx and Lucian in patch 4.20.
Additionally, buffed champions may be played at a higher rate, affecting the game balance. 
In the next section, we explore how patches can affect \emph{players} differently, rather than focus on the team performance with different champions.

\subsection{Individual Player Performance}\label{sec:player} 


By altering champion abilities and game settings, software patches can potentially affect player performance. Figure~\ref{fig:version_heat} shows average player performance per match, measured by the number of kills the player makes, for games played after each new patch was introduced. 
 The figure shows two views of the same data. The top heatmap disaggregates player matches by the value of the feature \textit{timeSinceLastMatch}, which measures time elapsed since the last game they played.  The top line reports the average performance of players who  move on to the next match without taking a break (\textit{timeSinceLastMatch}=0). These players generally perform poorly, as evidenced by the darker colors in the top line. This is consistent with the finding by Sapienza et al.~\citeyear{sapienza2018individual} that player performance deteriorates over the course of a gameplaying session due to cognitive depletion. This suggests that players become fatigued and their performance declines. The following lines report the corresponding percentiles of the remaining values of the \textit{timeSinceLastMatch} feature. Players in second line (26\%) taking a short break (< 3 minutes) generally outperform other players, including those who take longer breaks, from days (46\%) to years (82\%). This could potentially mean that  players taking short breaks are more dedicated and able to play more frequently, thereby improving in skill. Also interesting is the dark band across all bins that is seen before patch 6.9. This means that all groups of players---those taking short and long breaks---performed worse on average than in the earlier versions. 

The heatmap at the bottom of Figure~\ref{fig:version_heat} shows the same data 
but  disaggregated by \textit{meanKillsAtStart}. The top row (29\%) shows players with   \textit{meanKillsAtStart} equal to zero, and the remaining players split by quartiles of the feature value. This feature is a proxy of player skill: new players are grouped in the first bin, with remaining players grouped into bins from weakest players with few kills per match (29\%) to the strongest players with many kills per match (82\%). Unsurprisingly,  better players have better performance (last two lines are brightest) and outperform weaker players in all versions of the game. New players outperform  the weakest players, which makes sense, as new players have a range of skill. 

We see abrupt color shifts in the heatmaps in Figure~\ref{fig:version_heat}, especially between versions 4.19 and 4.20 and versions 6.8 and 6.9. These color shifts within a row indicate significant changes in player performance that come with new versions. Comparing performance of players before and after the version change allows for measuring the causal effect of the software patch on players.

\subsubsection{Average effect of patches}
First we estimate the overall effect averaged for all players. For every game version, we measure its average effect on performance by calculating the difference in the mean number of \textit{kills} before and after the version change.
Figure~\ref{fig:version_bar} shows mean effect on performance introduced over 62 software patches in our data. 
The patch version 4.20 introduces game changes  that increase the average number of \textit{kills} made by a player by 0.5 per match. This is a pre-season patch, and our results show that it has major impact on performance. 
In patches immediately following version 4.20, the effect is slightly negative---as if to compensate for the changes made in version 4.20. Version 6.9 is another major patch, which increases the average number of \textit{kills} by over 0.4. 
The remaining patches have a far smaller effect on average.


\begin{figure*}[!htb]
    \begin{minipage}{1.0\linewidth}
        \centering        
        Mean effect of patches on performance disaggregated by feature \textit{timeSinceLastMatch} \\
        \includegraphics[width=1.0\linewidth]{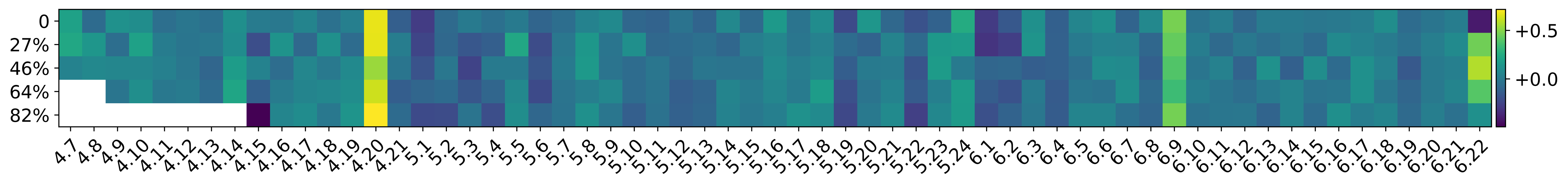}
    \end{minipage}
    \begin{minipage}{1.0\linewidth}
        \centering        Mean effect of patches on performance disaggregated by feature \textit{meanKillsAtStart} \\
        \includegraphics[width=1.0\linewidth]{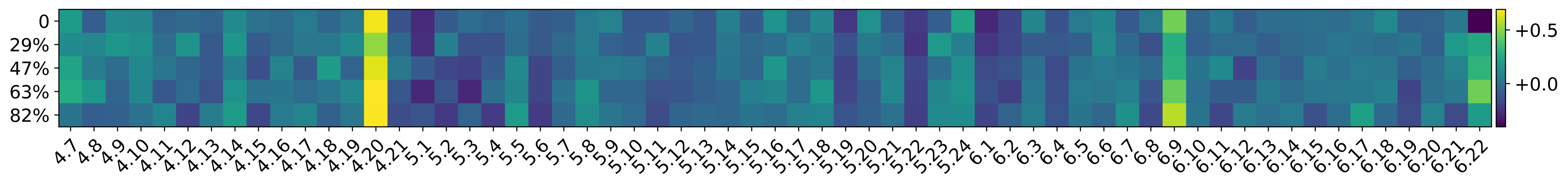}
    \end{minipage}
    \caption{Causal impact of game patches on player performance. 
    Heatmap shows the \textit{effect} of the software patch on performance for the same groups of players as in Figure~\protect\ref{fig:version_heat}. 
    The top figure shows mean effect for players with different rest time (\textit{timeSinceLastMatch}), and the bottom plot shows the effect of patches for bins with different values of \textit{meanKillsAtStart}. 
    }
    \label{fig:version_heat_effect}
\end{figure*}

\subsubsection{Heterogeneous effect of patches}
The mean effect hides much of the complexity of the impact of software changes on different players. 
We can see some of this heterogeneity 
in Figure~\ref{fig:version_heat_effect}, which compares the mean effect for  players disaggregated into the same groups as Figure~\ref{fig:version_heat}. Again, we see a strong effect of patches 4.20 and 6.9 on all groups of players, although some groups are better able to leverage changes made in the game than other players. This analysis, however, does not account for the impact of the champion (or champion type) the player chooses on performance. To address this point,  we learn a causal tree for each champion. 

To learn the causal trees from data, we first combine data from two consecutive versions of the game and then disaggregate the combined data by champion. We use the combined data to learn the causal tree for the champion and repeat for all pairs of consecutive game versions. 
Due to limited data and resources, we only select the top 25 most popular champions.  
We set minimum leaf node size as 5\% of total samples and maximum depth of causal tree as 10.



\begin{figure*}[!htb]
    \begin{minipage}{0.40\linewidth}
        \centering
        \includegraphics[width=1.0\linewidth]{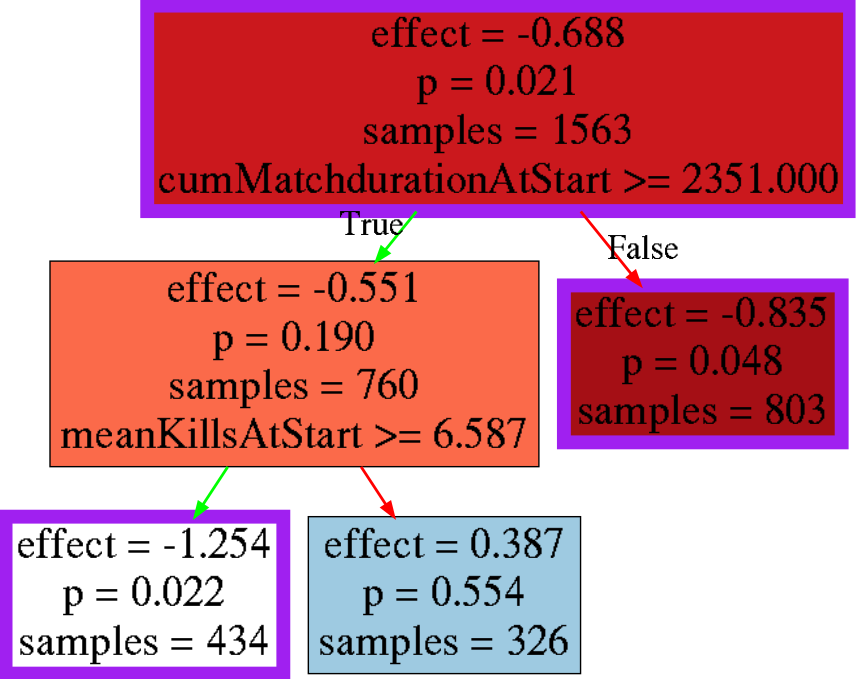}
    \end{minipage}
    \begin{minipage}{0.40\linewidth}
        \centering
        \includegraphics[width=1.0\linewidth]{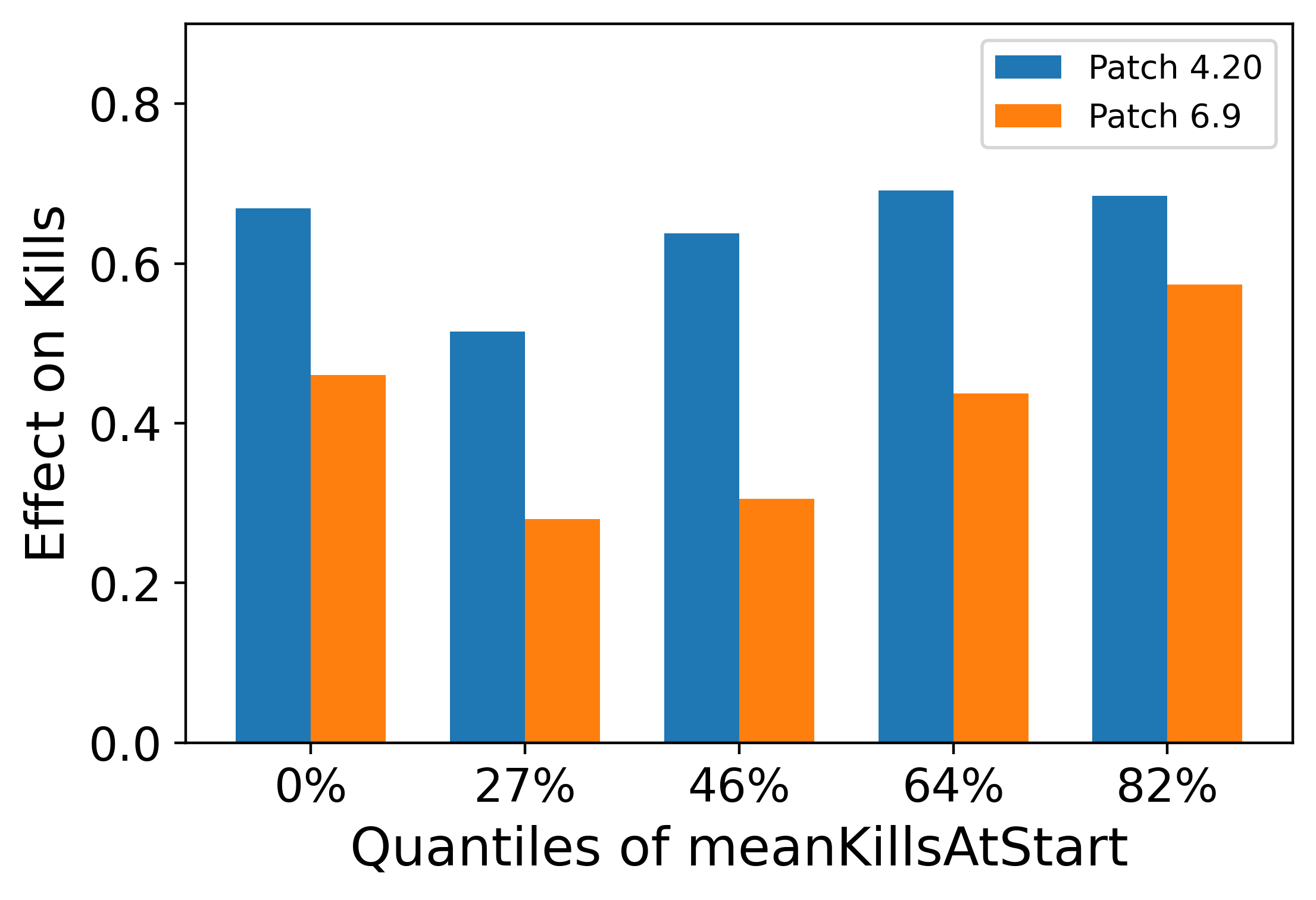}
    \end{minipage}
\caption{(Left). Causal tree learned for patch 4.21 for matches where a player select champion \textit{Vayne}. The purple nodes show heterogeneous effect that is significant at 5\% level. 
(Right). Average treatment effect calculated for players with different levels (\textit{meanKillsAtStart}) at two major patch changes. Excluding the first bin which contains significant portion of new players with \textit{meanKillsAtStart} close to zero, we see a trend that high level players benefit more from the patch changes, indication the performance gap being widen.
}
\label{fig:kills_tree}
\end{figure*}

Our causal modeling framework learns 1,550 causal trees for 25 champions over 62 software patches. Figure~\ref{fig:kills_tree} shows a pruned tree learned for \textit{Vayne} for versions 4.20-4.21. The causal tree identifies two leaf nodes to be statistically significant at 5\% level, which are highlighted in purple. The leaf node at the second level demonstrates that for a less experienced player (
cumulative match duration is small), the effect is lower than average effect (at the root node). Leaf nodes at the lowest level show that the treatment effect is smaller than average for experienced and good players (high value of  \textit{cumMatchdurationAtStart} and \textit{meanKillsAtStart}). 

\begin{figure}[!htb]
    
    \begin{minipage}{0.9\linewidth}
        \centering
        \includegraphics[width=\linewidth]{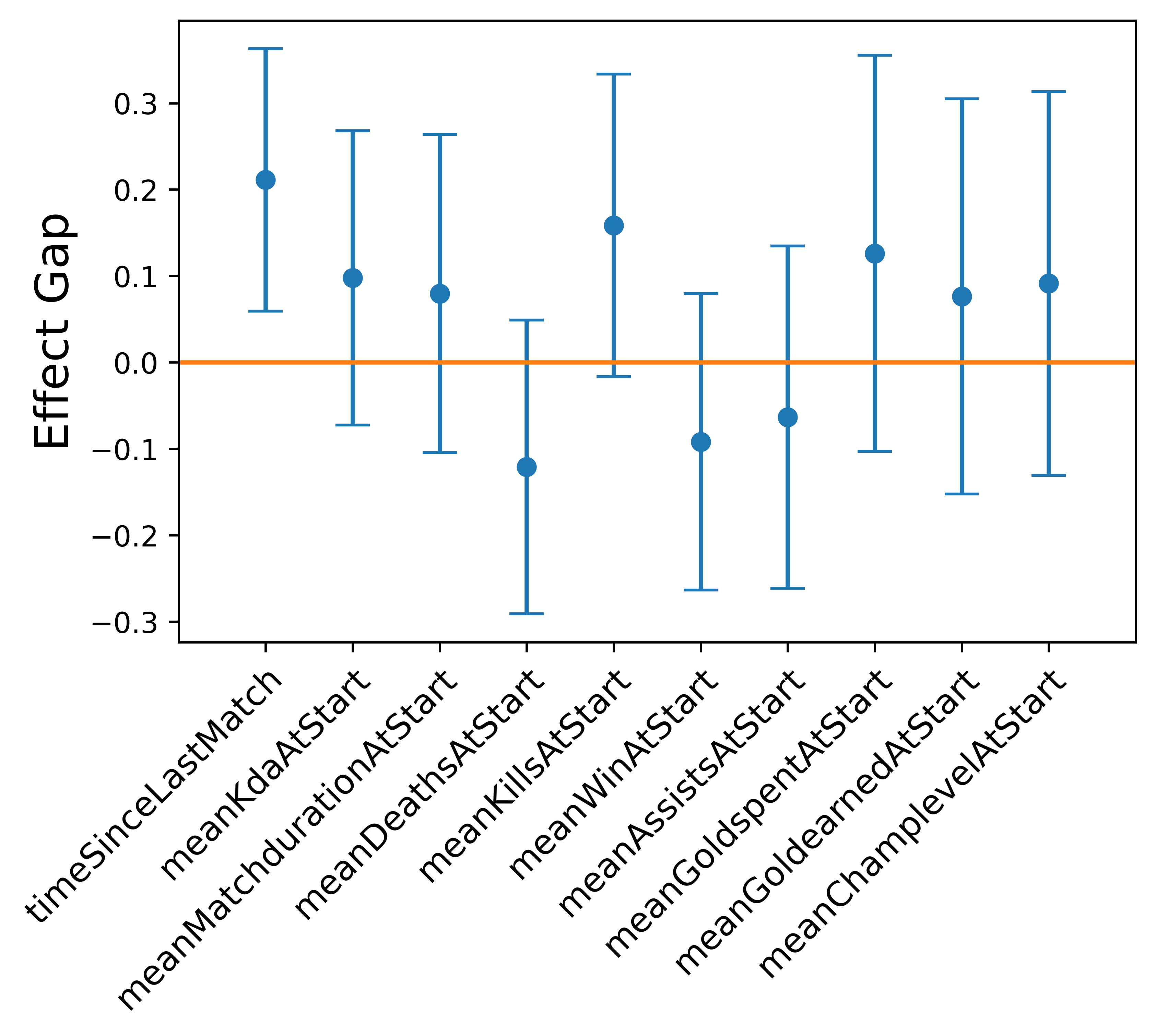}
    \end{minipage}
    \caption{
    Average effect gap calculated for the 10 most important features. The error bar shows 95\% confidential intervals.
    We can see that the most important features are \textit{timeSinceLaseMatch} and history performance of players.
    }
    \label{fig:feature_bar}
\end{figure}

\subsubsection{Effect of player features}
To explore the causal effects of player characteristics on their performance, we quantify the relationship between player features and the heterogeneous treatment effect of software patches. We perform statistical analysis of the features used to split nodes. We weigh each feature in tree by the sample size in the split and use the total weight (among all champions and versions) to compare the relative importance of the features. The important features are those that occur in many trees and explain many data points. 
The ten most important features related to changes in champions are (in descending order): \textit{timeSinceLastMatch}, \textit{meanKdaAtStart},
\textit{meanMatchDurationAtStart}, \textit{meanDeathsAtStart}, \textit{meanKillsAtStart}, , \textit{meanWinsAtStart}, \textit{meanAssistsAtStart}, \textit{meanGoldspentAtStart}, \textit{meanGoldearnedAtStart}, and \textit{meanChamplevelAtStart}. 
Except for \textit{timeSinceLastMatch} and \textit{meanMatchDuration}, many of the important features are performance related. Interestingly, mean values of features such as kills, deaths, gold earned and spent, are judged to be more important than their cumulative values. The cumulative values are larger for players who play more games, but mean are calculated per match, and therefore, better reflect player's skill. However, more important than skill is the length of the break between games.

To quantify how player features, such as \textit{timeSinceLastMatch}, allow players to leverage patches to improve their performance,
we find all splits on \textit{timeSinceLastMatch} 
and calculate the difference of the heterogeneous treatment effect between the left child node (feature  larger than or equal to cutoff) and right child node (feature smaller than cutoff) for each split. We use this causal effect difference---\textit{effect gap}---to describe the overall impact of the feature on causal effects. If the difference is larger than 0, then the feature improves player performance after a patch. 
The effect gaps of important features are shown in Figure~\ref{fig:feature_bar}. Players with high \textit{meanKillsAtStart} have a positive effect gap, meaning that their performance tends to improve following a software patch, while the performance of players with high \textit{meanDeathsAtStart} tends to decrease. 
However, the p-values for these effects are slightly higher than 0.05, which suggests that skilled players (with high kills or low deaths) can only take weak advantage of changes in champions made by the software patches. 
This may be a consequence of game balance: when one group of players has advantage in one game version, designers reduce their advantage in the next version. For example, in the bottom heatmap in Figure~\ref{fig:version_heat_effect}, the positive impact on performance of highly skilled players in patch version 4.20 (bright colors in bin\_5) is offset by relatively stronger decreases (darker colors) in their performance  in patch 4.21.

The only feature with a significant effect gap (at 5\% significance level) is \textit{timeSinceLastMatch}. 
The positive effect gap suggests that players taking longer breaks between games (or at least those who do not play without interruptions) are consistently able to improve their performance. These results highlight the importance of short-term changes in player behavior: cognitive fatigue following uninterrupted gameplay diminishes players of ability to leverage changes in champions.

\section{Conclusion}

Gameplay is rarely explored from the perspective of individualized performance. When we study the heterogeneous effect of patches, we discover significant changes in team and player performance depending on champions, 
player's track record, and length of breaks between matches.
On a team level, we found that changes to champions have an impact on all champions selected by the team, and not just the ones that were changed.
Significant changes are generally identified as important splits in the causal tree, while small changes are ignored or shown in lower-level nodes. 
At a player level, several performance metrics demonstrate significant benefits of patches for high-performing players rather than low-performing ones. These patches therefore counter-intuitively widen the gap between the high and low-performance players.
When we analysed individual player performance, we also found that there are two major version changes, 4.20 and 6.9 that have an outsized impact on performance. But these changes hide the heterogeneity in the impact. 
Using causal trees, we studied the heterogeneous effect of patches on players with the same champions and found that \textit{timeSinceLastMatch} and performance proxies are the most important factors. 
On the other hand, we find that there is little or no 
correlation between these factors on conditional average treatment effects.

Because these results are based on observational data, however, they suffer from the limitations of all causal models suffer. Specifically, we cannot guarantee that players are selected at random into the control (before a patch) and treatment (after the patch) conditions. On the contrary, previous work shows that some players behavior changes after a patch to reflect the known champion buffs and nerfs~\cite{wang2020research}. Not only is there a potential selection bias, however, there might be potential confoundings from features we do not know, and therefore cannot control for. While these results help us understand the impact patches have on player performance ``in the wild'', future work should create controlled experiments to address these potential confounders. 

\subsection*{Acknowledgments}
This work was supported in part by DARPA under contract HR00111990114

\balance
\bibliographystyle{ACM-Reference-Format}
\bibliography{references}

\end{document}